\documentclass[aps,prb,reprint]{revtex4-1}
\usepackage[dvipdfmx]{graphicx}
\usepackage{url}
\usepackage{dcolumn}
\usepackage{bm}
\usepackage{epsfig}
\usepackage{braket}
\usepackage{amsmath}
\usepackage{here}   
\usepackage{amsmath}
\usepackage{amsfonts}   
\usepackage{xcolor}
\newcommand{\ch}[1]{{\color{black} #1}}

\begin{document}
\draft
\title{Designing Rashba-Dresselhaus effect in magnetic insulators}
\author{Masataka Kawano$^1$ , Yoshinori Onose$^{1,2}$ , and Chisa Hotta$^1$ }
\address{$^1$ Department of Basic Science, University of Tokyo, Tokyo, 153-8902, Japan}
\address{$^2$ Institute for Materials Research, Tohoku University, Sendai, 980-8577, Japan}
\begin{abstract}
One of the major strategies to control magnetism in spintronics is to utilize 
the coupling between electron spin and its orbital motion. 
The Rashba and Dresselhaus spin-orbit couplings induce magnetic textures of band electrons called spin momentum locking, 
which produces a spin torque by the injection of electric current. 
However, joule heating had been a bottleneck for device applications. 
Here, we propose a theory to generate further rich spin textures 
in insulating antiferromagnets with broken spatial inversion symmetry (SIS), 
which is easily controlled by a small magnetic field. 
In antiferromagnets, the ordered moments host two species of magnons 
that serve as internal degrees of freedom in analogy with electron spins. 
The Dzyaloshinskii-Moriya interaction introduced by the SIS breaking couples 
the two-magnon-degrees of freedom with the magnon momentum. 
We present a systematic way to design such texture and to detect it via magnonic spin current 
for the realization of antiferromagnetic memory.
\end{abstract}
\maketitle
\narrowtext
Development of tunable magnetic structure has long been a key issue 
for detecting and controlling magnetic domains electrically toward application to memory devices\cite{review}. 
Besides the conventional domain walls that appear in real space, 
particular focus is given on emergent spin textures in reciprocal space, called ``spin momentum locking". 
The spin textures are classified into 
Rashba-\cite{rashba1960,casella1960,bychkov1984} and Dresselhaus-types\cite{dresselhaus1955} 
that exhibit vortex- and anti-vortex geometries along the closed Fermi surfaces. 
Since the wave number $\bm k$ distinguishes the electronic state of matter, 
such spin texture allows for the selection of magnetic moment the state/current carries. 
This has brought about fundamentally important and technologically promising phenomena 
including spin Hall effect\cite{murakami03,kato04,sinova04,wunderlich05}, 
spin-orbit torque\cite{miron10,kurebayashi14}, and Rashba-Edelstein effect\cite{edelstein1990,kato2004,silov2004}. 
\par
In insulating magnets, an excitation is carried by the quasiparticle called magnon, 
which represents a quantum mechanical spin precession propagating in space. 
Such propagation is predominantly mediated by the standard magnetic exchange interaction $J \bm S_i \cdot \bm S_j$, 
between spins, $\bm S_i$ and $\bm S_j$. 
In a uniform ferromagnet, a simple exchange interaction, $J(<0)$, generates {\it non-degenerate} quadratic magnon bands. 
When \ch{the spatial inversion symmetry (SIS)} is broken, an antisymmetric spin exchange called Dzyaloshinskii-Moriya (DM) interaction\cite{dzyaloshinsky1958,moriya1960}, 
$\bm D\cdot (\bm S_i\times \bm S_j)$, appears, depending on the crystal symmetry. 
This term bends the propagation of magnons in space 
in a similar manner to the cyclotron motion of electrons in the presence of magnetic flux\cite{matsumoto11}. 
Thus, when $\bm D$ is parallel to the magnetization, magnon bands in a ferromagnet become asymmetric, 
reflecting the ``nonreciprocal" propagation\cite{melcher1973,kataoka1987,zakeri10,kaidi15,stashkevich15,hans15,sato2016,iguchi15}. 
Nevertheless, the phenomena related to magnons in non-centrosymmetric ferro or ferrimagnets lacks abundance 
compared to the rich counterparts of the conducting Rashba electrons. 
\ch{
This is because the ferromagnetic magnons carry spins that are pointing in a unique direction, 
and cannot afford up and down spin degrees of freedom like electrons. 
}
\par
In antiferromagnets ($J>0$) with a doubled magnetic unit cell, 
the magnon bands are folded and a {\it doubly-degenerate} linear dispersion appears at $\Gamma$-point. 
What if we regard two different species of magnons each belonging to the degenerate band as 
an analogue of the electronic spin degrees of freedom? 
If this degrees of freedom couples to the magnon momentum via a DM interaction, 
such that the SO does in electron systems, 
one may expect as rich phenomena as those of the Rashba-electronic systems in insulators. 
So far, however, non-descript bipartite antiferromagnets did not show any magnon-based phenomena 
such as a thermal Hall effect already found in the kagome and pyrochlore magnets
\cite{katsura10,onose12,onose10,matsumoto14,lee15,owerre17,hirschberger2015,hirschberger2015-2},
indicating that the situation is not as simple. 
In this article, we demonstrate that a typical two-dimensional (2D) antiferromagnet 
can afford a variety of spin textures not even found in electronic systems. 
We introduce a pseudo-spin degrees of freedom 
based on the two species of magnons belonging to antiferromagnetic sublattices, 
and show that the reduction of $SU(2)$ symmetry of these pseudo-spins is required to have 
spin textures in momentum space.
We classify the way how the symmetry is reduced step-by-step 
by the interplay of DM interaction, spin anisotropy, and magnetic field in a series of 1D antiferromagnets, 
which determines the degree of variety of the spin textures. 
A systematic way to design 2D spin textures is thus given based on the building blocks 
of these 1D antiferromagnets. 
\vspace{5mm}
\par
%
\begin{figure*}[t]
\includegraphics[width=18cm]{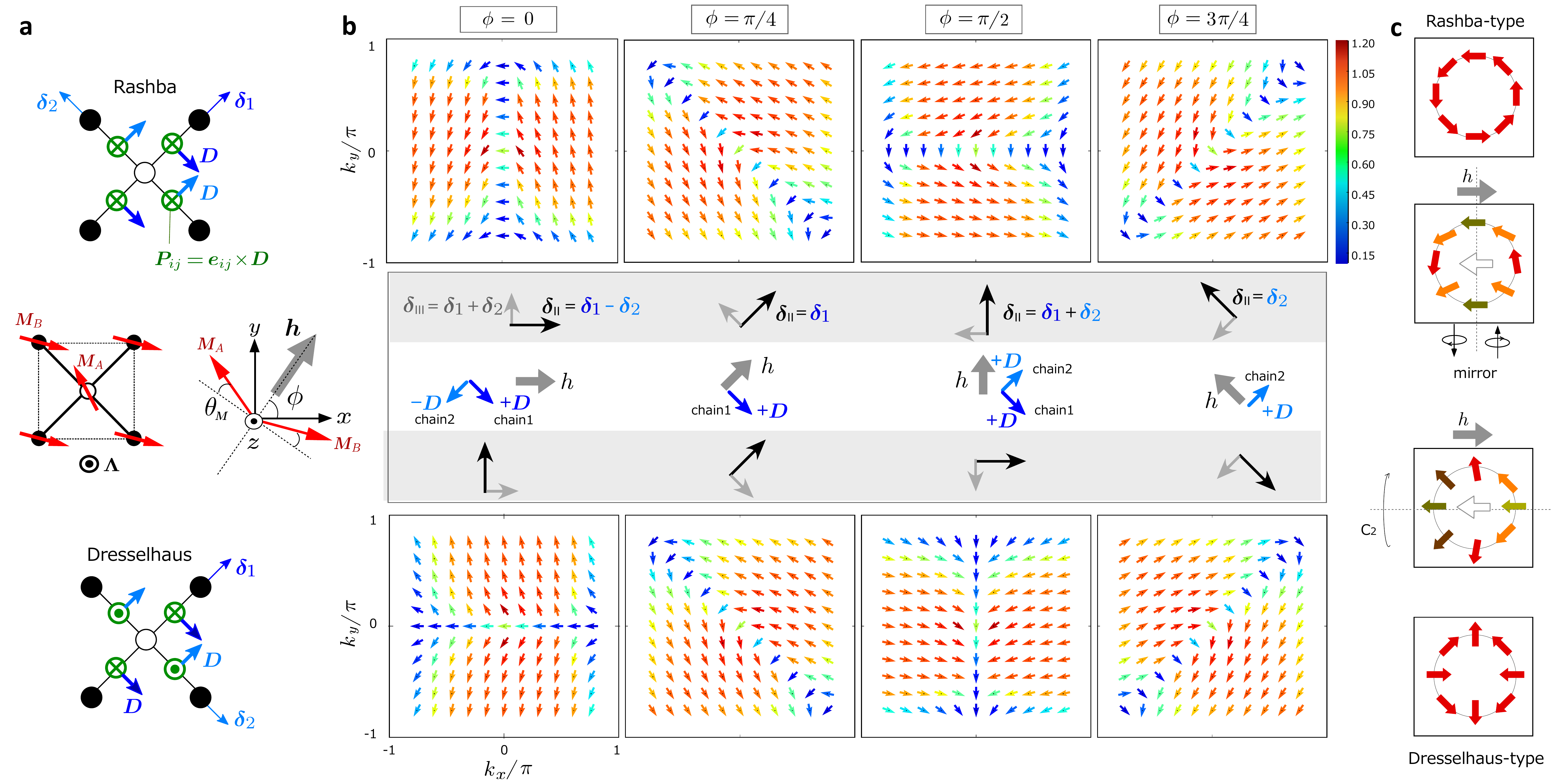}
\caption{
{\bf (a)} 2D SIS broken antiferromagnets with easy-$xy$-plane anisotropy ($\Lambda>0$). 
Open/filled circles represent the A/B magnetic sublattice, 
and the red arrows are the \ch{schematic} classical spin configurations $\bm M_A$ and $\bm M_B$ 
of the ground state \ch{characterized by the field angle $\phi$ and the canting angle $\theta_M$. }
The DM vector, $\bm D_{ij}$, has in-plane elements, 
depicted by defining $i\rightarrow j$ direction as unit vectors 
$\bm e_{ij}=+\bm{\delta}_{1}$ and $+\bm{\delta}_{2}$, for two bond directions. 
Top and bottom panels are Rashba/Dresselhaus-type configuration of $\bm D$, characterized by the 
polarization vector $\bm P_{ij}=\bm e_{ij}\times \bm D$. 
{\bf (b)} Textures of $\bm{S}_{\bm{k},-}$ of the lower magnon band $\omega_-(\bm k)$ 
over the first Brillouin zone with $S=5/2$, $J=1.0$, $D=0.1$, $\Lambda=0.05$, and $h=3J$. 
Upper and lower panels are those of the Rashba- and the Dresselhaus-type ones for 
field angles, $\phi=0,\pi/4,\pi/2$, and $3\pi/4$. 
The amplitude, $|\bm{S}_{\bm{k},-}|$, is given by the density plot, 
and the arrow indicates its direction. 
The middle panels show how the direction with (long black arrow: $\bm \delta_{\rm II}$) 
and without(short gray arrow: $\bm \delta_{\rm III}$) spin textures (see Fig.\ref{f3}{\bf d}) 
is elucidated by the relative relationships between the summation of $\bm D$ 
of the two different bond directions and $\bm h$. 
{\bf (c)}  Sketch of how the typical Rashba-type and Dresselhaus-type spin textures 
are modified according to the magnetic field and to the symmetries kept at $\phi=0$.}
\label{f1}
\end{figure*}

\ \\
\noindent
{\bf Results}\\ 
\noindent
{\bf \small Model systems}
\\
We consider a quantum spin system with nearest neighbor exchange interaction, $J$($>0$), 
spin anisotropy, $\Lambda$, the DM interaction, $\bm D_{ij}$, 
and a uniform external magnetic field, $\bm h$. 
The general form of the Hamiltonian is given as 
\begin{eqnarray}
{\cal H}&=&J\sum_{\langle i,j\rangle}\bm{S}_{i}\cdot\bm{S}_{j}
+\sum_{\langle i,j\rangle}\bm{D}_{i,j}\cdot[\bm{S}_{i}\times\bm{S}_{j}] \nonumber \\
&&+ \Lambda\sum_{i} (S_{i}^{\Lambda})^{2}-\sum_{i}\bm{S}_{i}\cdot\bm{h}, 
\label{ham}
\end{eqnarray}
where $S_{i}^{\Lambda}$ is the spin moment in the $\bm \Lambda$-direction. 
\ch{In our 2D antiferromagnet, we consider an easy plane anisotropy normal to $\bm \Lambda$-axis ($\parallel z$), 
namely $\Lambda>0$. 
If we take $\Lambda<0$, an easy axis anisotropy is realized, which we apply shortly to a series of 1D magnets. 
We take the $x$- and $y$-axes in the direction rotated by $\pi/4$ from the bond direction. 
We place the in-plane magnetic field $\bm h$ perpendicular to the $z$-axis.}
The DM interaction emerges when the midpoint of two magnetic sites lacks the inversion symmetry. 
There are two different ways of aligning the $\bm D$-vector; 
In defining the spin indices that couples to $\bm D_{ij}$ in an order, $i \rightarrow j$, 
the $\bm D$-vectors can take either a uniform or a staggered 
configuration along that direction. The former breaks 
the global SIS of the crystal, whereas the latter keeps
the site-centered SIS. Since the staggered DM interaction does not play any role in the physics presented here,
we focus on the uniform DM in the following.
\par
We consider the physics realized for small $D_{i,j}$ and $\Lambda$ compared to $J=1$, 
\ch{where we always find a canted antiferromagnetic order in the ground state. 
The spin moments, $\bm M_A$ and $\bm M_B$, on the magnetic two-sublattices 
are confined within the easy plane, 
and their directions are described by the in-plane magnetic field angle, $\phi$ against the $x$-axis, 
and the canting angle, $\theta_M$ (see Fig.\ref{f1}{\bf a}). 
The degree of spin canting is as small as 
$\theta_M=\arcsin(h/8JS)\sim 1.5^\circ$ and $8^\circ$
for $h=0.5J$ and $3J$, respectively. }
The magnetic excitations are described by two species of Holstein-Primakoff bosons (magnons)\cite{holstein40}; 
The $z$-component of spin operators belonging to sublattices A an B is given as 
$S_i^z= S - a^\dagger_i a_i$ and $S_i^z=-S + b^\dagger_i b_i$, 
where $a^\dagger_i/a_i$  and $b^\dagger_i/b_i$, are the magnon creation/annihilation operators. 
The resultant spin wave Hamiltonian of magnons spanned on a particle-hole space is given as 
\begin{equation}
{\cal H}_{\rm SW} = \frac{1}{2}\sum_{\bm{k}}\Phi_{\bm{k}}^{\dagger}H_{\mathrm{BdG}}(\bm{k})\Phi_{\bm{k}}, 
\label{eq:uhu}
\end{equation}
where $H_{\mathrm{BdG}}(\bm{k})$ takes the form of the bosonic Bogoliubov-de Gennes (BdG) Hamiltonian (see Methods), 
and $\Phi_{\bm{k}}=(a_{\bm{k}},b_{\bm{k}},a^{\dagger}_{-\bm{k}},b^{\dagger}_{-\bm{k}})^{T}$ 
represents the particle and hole pairs of A and B magnons in reciprocal space(see Methods). 
The eigenvalues of $H_{\mathrm{BdG}}(\bm{k})$ give the magnon bands, $\omega_\pm (\bm k)$. 
The spin moment carried by magnons on these bands at each $\bm{k}$ is evaluated as\cite{okuma2017},
\begin{equation}
\bm{S}_{\bm{k},\pm}=-\bm{M}_{\mathrm{A}}d_{\mathrm{A}}(\omega_{\pm}(\bm{k}))-\bm{M}_{\mathrm{B}}d_{\mathrm{B}}(\omega_{\pm}(\bm{k}))
\label{Smag}
\end{equation}
using the local spectral weight at $\bm{k}$ of $\omega_{\pm}$-band, $d_{\mathrm{A}/\mathrm{B}}(\omega_{\pm}(\bm{k}))$, on sublattice A/B (see Methods).
\vspace{5mm}
\\
\noindent
{\bf \small Rashba and Dresselhaus magnons}
\\
We first demonstrate that the 2D SIS broken antiferromagnets can exhibit as rich spin textures as 
those of Rashba and Dresselhaus electronic systems.  
For a 2D square lattice, there are two different ways of constructing a model Hamiltonian Eq.(\ref{ham}) 
in terms of uniform DM interaction; 
We call the ones shown in the upper and lower panels of Fig.\ref{f1}{\bf a} 
Rashba and Dresselhaus antiferromagnets, respectively. 
Here, $\bm D_{ij}$ vectors are depicted in the directions defined by taking the indices $i\rightarrow j$ pointing in the 
$+\bm \delta_1$ and $+\bm \delta_2$ directions along the two bonds.
We do not consider explicitly the component of $\bm D_{ij}$ normal to the 2D plane. 
This is because the out-of-plane component generally takes the form of the staggered DM interaction 
as can be elucidated for the case of Ba$_2$MnGe$_2$O$_7$ with space group $P\bar{4}2_1m$\cite{murakawa2012}, 
which contributes neither to the spin textures nor to the nonreciprocity of magnon bands. 
(see Supplementary material D for details). 
\par
It is convenient to classify the two types of antiferromagnets by a polarization vector, $\bm P_{ij}= \bm e_{ij} \times \bm D$, 
where the unit vector $\bm e_{ij}$ points to either $+\bm \delta_1$ or $+\bm \delta_2$ along the bonds. 
The Rashba-type antiferromagnet in the upper panel of Fig.\ref{f1}{\bf a} has the $C_{4}$ symmetry 
where bulk polarization is induced by the broken SIS.
This kind of polarization is equivalent in the symmetry to the ones induced by the field perpendicular 
to the plane, whose gradient generates a Rashba SO coupling.
For the one in the lower panel of Fig.\ref{f1}{\bf a},
the DM vector keeps the $\bar{C}_{4}$ symmetry, 
where bulk polarization is absent even though the global SIS is broken.
It is realized in crystals with $D_{2d}$ or $T_{d}$ point group symmetry, 
and is related to the Dresselhaus-type of SO interaction. 
Another way to look at is to perform a $C_4$-rotation to $\bm \delta_1$ and $\bm \delta_2$ 
in the clockwise direction, and we find that $\bm D$ rotates in the anti-clockwise direction 
in the Dresselhaus-case, and clockwise in the Rashba-case. 
\par
\begin{figure}[t]
\includegraphics[width=9cm]{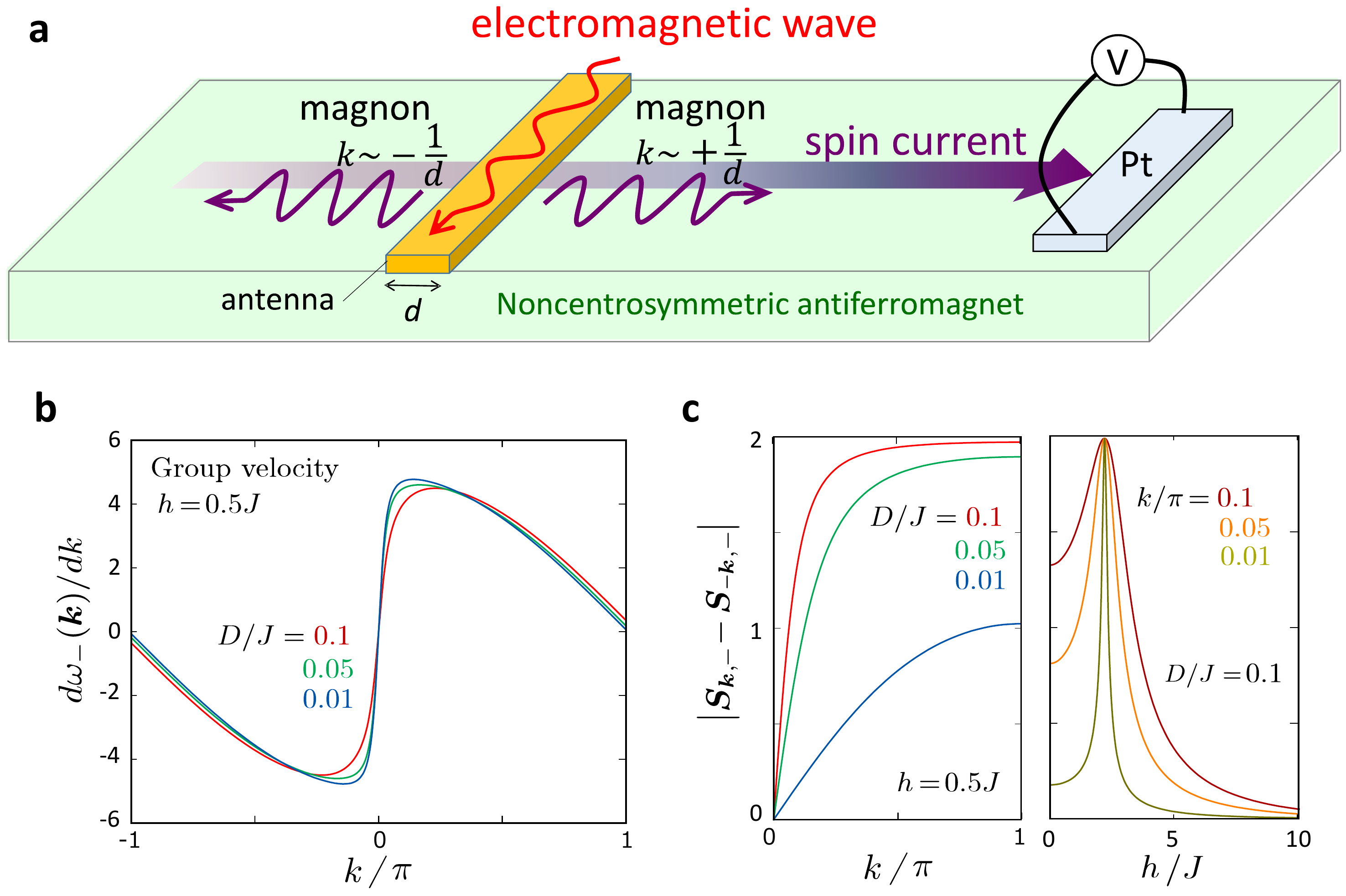}
\caption{
{\bf (a)} Illustration of the experiment to probe the structure of a noncentrosymmetric antiferromagnet by 
generating a nearly pure spin current of wave numbers, $\bm k$ and $-\bm k$. 
The width of the antenna is designed to selectively excite magnons of $|\bm k|\sim 1/d$. 
The inverse spin Hall voltage of Pt electrode is used to detect the spin current.
\ch{
{\bf (b)} Group velocity $d\omega_-(\bm k)/d k$ 
and (c) $|\bm S_{\bm k,-}-\bm S_{-\bm k,-}|$, 
for $k=k_x$ ($k_y=0$ line for Rashba-type $\phi=0$ in Fig.1{\bf b}) 
with $D/J=0.1, 0.05$ and 0.01, where we set $h=0.5J$. 
The right panel is the $h$-dependence for $k/\pi=0.1,0.05$ and 0.01. 
At $h\sim 2J$, the two magnon bands touch at $\Gamma$-point, which enhances 
the mixing of spins and gives a sharp peak in 
$|\bm S_{\bm k,-}-\bm S_{-\bm k,-}|$. }
}
\label{f4}
\end{figure}

Figure \ref{f1}{\bf b} shows the direction and the amplitude of the spin carried by magnons, $\bm{S}_{\bm{k},\pm}$, 
of the lower magnon band, $\omega_-(\bm k)$ over the Brillouin zone for the Rashba and Dresselhaus magnons. 
We take $D/J=0.1$, $\Lambda/J=0.05$, and $h/J=3.0$. 
The upper band not shown here hosts similar texture, while the component perpendicular to $\bm h$ 
points in the opposite direction
(see Supplementary material Fig.S{\bf 2}). 
The total net moment is opposite to $\bm{h}$. 
This is because the magnons represent the shrinking of classical magnetization by definition, 
and the magnetic moments cant toward $\bm h$. 
A series of panels show how the spin textures evolve by the rotation of the magnetic field. 
\ch{The magnitude of $\bm h$ controls the amplitude of $\bm{S}_{\bm{k},\pm}$, 
while the same order of $|\bm{S}_{\bm{k},\pm}|$ and as rich texture sustain down to $h\rightarrow 0$ 
(see Fig.\ref{f4}(a) Supplementary Fig.S3)}. 
\par
One can understand overall spin textures in analogy with the well-known electronic counterparts.
The top and bottom panels of Fig.\ref{f1}{\bf c} show 
the typical Rashba-type and Dresselhaus-type spin textures observed in electronic systems; 
the former has $C_4$ symmetry about the $k_z$-axis and four mirror planes
$(k_x,k_y)=(0,1)(1,0)$, $(1,\pm 1)$, 
and the latter has $\bar{C_4}$ about the $k_z$-axis, $(1,\pm 1)$-mirror planes, 
and $C_2$ symmetries about the $k_x$- and $k_y$-axes. 
Note that the Hamiltonian of the Rashba-type and Dresselhaus-type antiferromagnets 
shown in Fig.\ref{f1}{\bf a} have the same crystal point group symmetry with 
their electronic counterparts. 
After modifying them to have a net moment opposite to the magnetic field, $\bm{h}$,
we see a rough orientation of $\bm{S}_{\bm{k},\pm}$ over the $\bm k$-space; 
the examples are given in the middle two panels of 
Fig \ref{f1}{\bf c} for $\bm{h}\parallel\bm{e}_{x}$ ($\phi=0$). 
The magnetic ordering and the magnetic field break the point group symmetry, 
but some of the space group symmetries are preserved, 
e.g. the Rashba-type antiferromagnet keeps a glide symmetry along the $y$-axis, 
so that the spin textures are invariant under mirror operation along the $k_y$-axis. 
The Dresselhaus-type one keeps two-fold screw axis along the $x$-axis, 
and thus the textures remain unchanged after the $C_{2}$ operation about the $k_x$-axis. 
\vspace{5mm}
\\
{\bf \small Types of spin textures} 
\\
Even though the spin momentum locking itself refer to the fixing of the angle between particle momentum $\bm k$ and spin moment $\bm{S}_{\bm{k},\pm}$, 
it does not necessarily mean the emergent texture of spins. 
To be more precise, there are three classes of spin textures, 
(i) spin moment is quantized in each band, 
\ch{namely $|\bm S_{\bm k,\pm}|=1$ and the direction is unique throughout each band}, 
(ii) amplitude of the spin moment depends on $\bm k$, but the direction is unique for each band, 
and (iii) both the direction and amplitude of the spin moment vary with $\bm k$.
Among them, only (iii) affords rich spin texture as in Fig.\ref{f1}. 
Ref.[\onlinecite{okuma2017}] showed the sufficient condition to have (i) 
(see Supplementary material B). 
\par
One can afford (i) and (ii) even when $\bm D=0$. 
To have case (iii) the uniform DM interaction is thus important. 
For example, if we take $\bm D=0$ in Eq.(\ref{ham}), we find case (ii), with $d_{\mathrm{A}}(\omega_{\pm}(\bm{k}))=d_{\mathrm{B}}(\omega_{\pm}(\bm{k}))=d(\omega_{\pm}(\bm{k}))$ and
\begin{equation}
\bm{S}_{\bm{k},\pm}=-(\bm{M}_{\mathrm{A}}+\bm{M}_{\mathrm{B}})d(\omega_\pm(\bm k)).
\label{eq:uni-S}
\end{equation}
As we see shortly, {\it the uniform DM interaction and the noncollinear spin texture 
together serve as a necessary condition for (iii) in antiferromagnets described by Eq.(\ref{ham})}.
The components of the magnetic field and the staggered DM interactions perpendicular 
to the plane do not contribute. 
\begin{figure}[t]
\includegraphics[width=8.5cm]{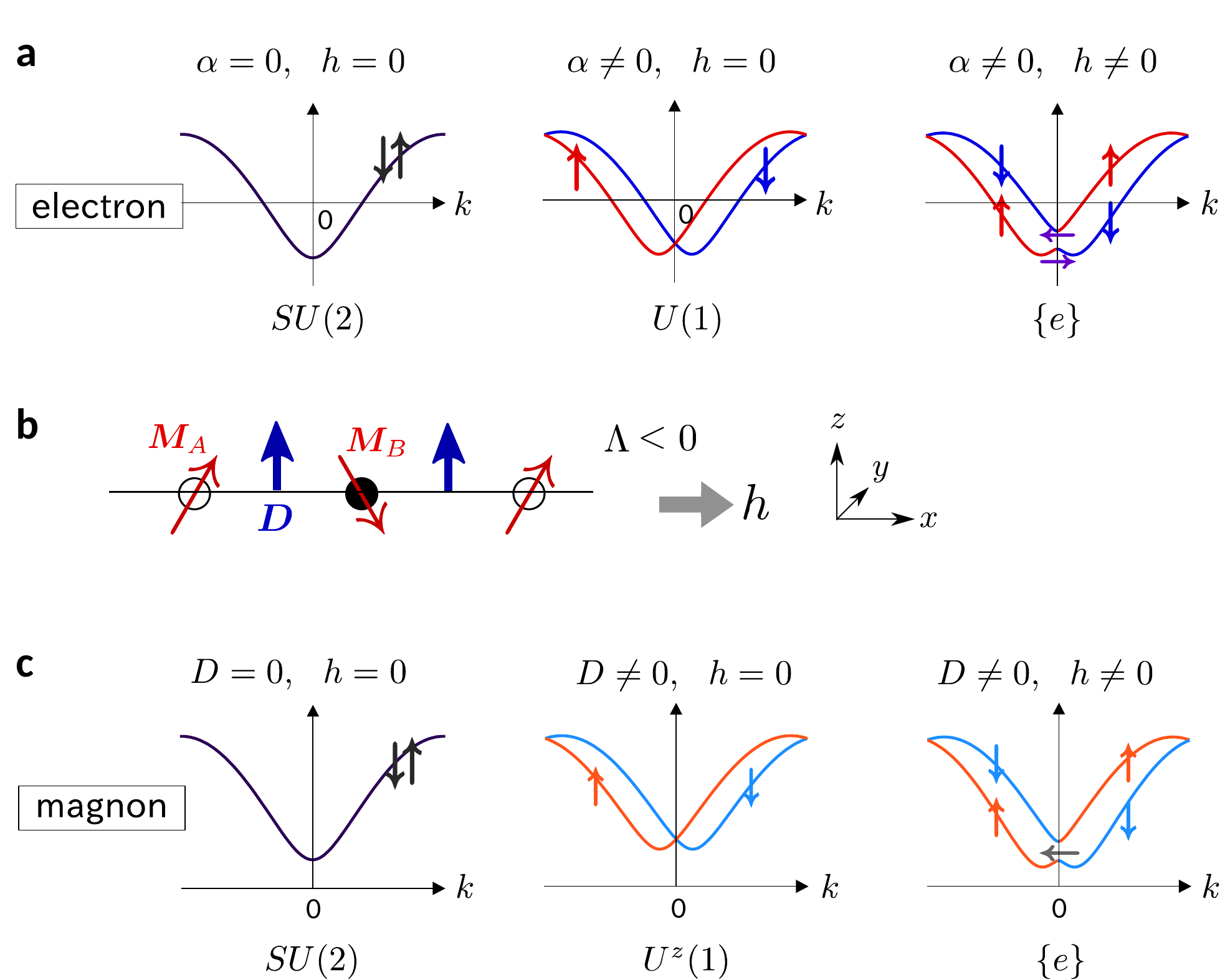}
\caption
{{\bf (a)} Dispersion of the 1D electronic system with the spin-orbit coupling and the magnetic field.
{\bf (b)} The 1D antiferromagnet with the uniform DM interaction and the magnetic field.
Red arrows indicate the classical spin configuration in the ground state,
and blue arrows indicate the uniform DM vector.
{\bf (c)} Dispersion of the 1D antiferromagnet with the uniform DM interaction and the magnetic field.
The DM interaction in the antiferromagnet corresponds to the spin-orbit coupling in the electronic system.
}
\label{f2}
\end{figure}
\begin{figure*}[t]
\includegraphics[width=18cm]{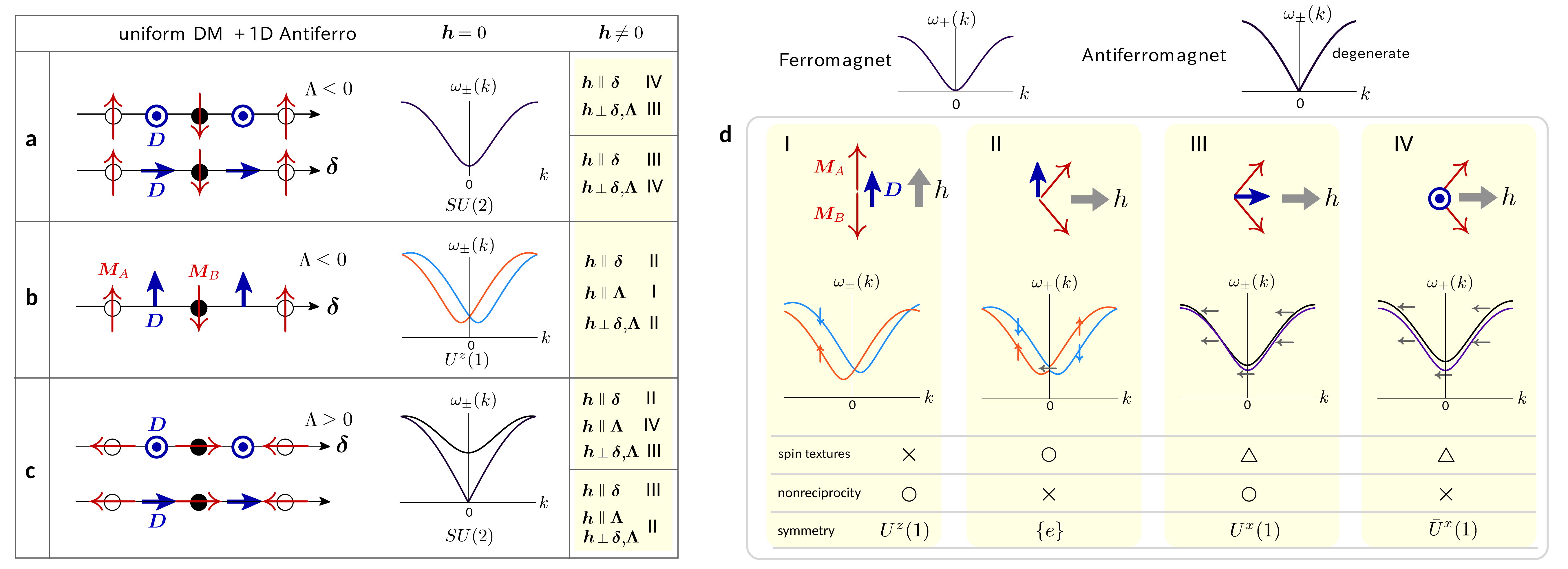}
\caption{
{\bf (a)-(c)} Schematic illustrations of representative 1D antiferromagnets with magnetic anisotropy and uniform DM interaction.
Magnon bands calculated for $\bm{h}=\bm{0}$ and $\bm{D}\neq\bm{0}$ are shown. 
{\bf (d)} Relative relationships between $\bm M_A$, $\bm M_B$, $\bm D$, and $\bm h$, classified into four cases I-IV. 
All possible types of 1D antiferromagnets including (a)-(c) fit one of these cases when $\bm h\ne 0$. 
Among three different classes of spin textures given in the main text, 
symbol $\times$ denotes (i) having quantized spin moment with $U(1)$ symmetry about the $z$-axis, 
$\triangle$ denotes (ii) $\bm k$-dependent $|S_{\bm k,\pm}|$ with $U(1)$ symmetry about the $x$-axis, 
and $\bigcirc$ a fully $\bm k$ dependent spin texture (iii) with broken symmetry, $\{e\}$. 
The nonreciprocity is present for (i) and (ii) with $\bigcirc$ and absent for the other two. 
}
\label{f3}
\end{figure*}
\vspace{5mm}
\\
{\bf \small Analogy of two-sublattice magnons with electrons }
\\ 
We see how one can qualify the antiferromagnets a property analogous to the electronic systems with SO coupling. 
(Details of the formulation are given in Supplementary material A. )
To this end, we consider a noninteracting 1D electronic system with a SO coupling, $\alpha$, and a magnetic field, $h$, 
shown in Fig.\ref{f2}{\bf a}. 
The Hamiltonian is given as ${\cal H}=(-2t\mathrm{cos}k)\sigma^{0}+(2\alpha\mathrm{sin}k)\sigma^{z}-h\sigma^{x}$, 
where $\sigma^{\mu}$ ($\mu=x,y,z$) is the Pauli matrix 
whose $z$-component classifies the up and down spin-1/2 of electrons, 
and $\sigma^{0}$ is a unit matrix. 
When $\alpha=h=0$, the energy bands are doubly degenerate as shown in Fig.\ref{f2}{\bf a}, 
which implies an $SU(2)$ symmetry represented by the operator $\sigma^\mu$. 
By the introduction of $\alpha\neq 0$, the two bands split, and the symmetry reduces from $SU(2)$ to $U(1)$. 
Because of this $U(1)$ symmetry, the spin moments the two bands carry are allowed to point only along the $z$-direction up and down. 
Finally, the magnetic field $h\ne 0$ breaks the $U(1)$ symmetry down to $\{e\}$, 
and the magnetic moments the energy bands carry start to depend on $\bm k$. 
The extension of the above discussion to 2D is straightforward: 
to realize Rashba or Dresselhaus types of spin textures, the symmetry reduction of $SU(2)$ to $\{e\}$ is required. 
\par
A similar argument applies to antiferromagnets. 
Let us consider a 1D system with a DM interaction and a magnetic field shown in Fig.\ref{f2}{\bf b}, 
whose Hamiltonian is given as Eq.(\ref{ham}) 
with $\bm{D}_{j,j+1}=D\bm{e}_{z}$, $\bm{\Lambda}=\Lambda\bm{e}_{z}$ ($\Lambda<0$), and $\bm{h}=h\bm{e}_{x}$. 
For the simplest antiferromagnet with $\Lambda=D=h=0$, the corresponding BdG Hamiltonian takes the form, 
$H_{\mathrm{BdG}}^{SU(2)}=(\tau^{0}\otimes\sigma^{0}) 2JS + (\tau^{x}\otimes\sigma^{x}) 2JS \cos (k)$, 
where $\tau^{\mu}$ and $\sigma^{\mu}$ ($\mu=x,y,z$) are Pauli matrices acting on a particle-hole space and a sublattice space, respectively. 
It has an $SU(2)$ symmetry, to which we intend {\it not in terms of the spin operator, 
but of the product space of sublattice and particle-hole degrees of freedom. }
The $SU(2)$ operator to classify this symmetry is given by 
\begin{equation}
J^{x}=\tau^{0}\otimes\sigma^{x},
\hspace{10pt}
J^{y}=\tau^{z}\otimes\sigma^{y},
\hspace{10pt}
J^{z}=\tau^{z}\otimes\sigma^{z}, 
\label{eq:p-SU2}
\end{equation}
which all commute with $H_{\mathrm{BdG}}^{SU(2)}(\bm{k})$. 
The reduction of $SU(2)$ to $U^z(1)$ symmetry is done by introducing $D \ne 0$ supported by $\Lambda>0$, 
$H_{\mathrm{BdG}}^{U^z(1)}=(\tau^{0}\otimes\sigma^{0}) 2(J+\Lambda)S + (\tau^{x}\otimes\sigma^{x}) 2JS \cos (k)
-(\tau^{y}\otimes\sigma^{y})2DS \sin(k)$. 
Among $J^\mu$'s, only the $z$-component fulfills $[H_{\mathrm{BdG}}^{U^z(1)}(\bm{k}),J^{z}]=0$. 
Here, we specify $U^z(1)$ as the $U(1)$ symmetry about the $z$-axis 
in order to discriminate from the one about the $x$-axis we see shortly. 
The classification of $U(1)$ about the spatial axis is required, 
since the definition of Pauli matrices $\tau^\mu$ are sticked to the real space axis, 
which is not the case for the usual electronic spins. 
\par
To obtain a $\bm k$-dependent spin texture, 
one further needs to break the $U^z(1)$ symmetry of magnons, 
which is done by applying the magnetic field in the $x$-direction. 
Here, the magnetic field parallel to $z$ only adds to the Hamiltonian a $(\tau^0\otimes \sigma^z)$ term which does not break the $U^z(1)$ symmetry. 
\par
Notice that the magnon spin textures near $k=0$ differ from those of the electronic systems. 
Also, the energy dispersions differ in that the electrons have Fermi level, 
whereas the magnons do not, and instead have a particle-hole gap. 
Although the particle-hole symmetric form of the BdG Hamiltonian makes the formulation rather complicated, 
the A and B sublattice degrees of freedom thus plays a role similar to the electronic spins, 
and so as $D\ne 0$ to the SO coupling of electrons. 
The full breaking of $SU(2)$ symmetry down to $\{e\}$ gives a necessary and sufficient condition 
to afford $\bm k$-dependent spin texture in 2D as well as in 1D antiferromagnets. 
\vspace{5mm}
\\
\noindent
{\bf \small 1D antiferromagnets as building blocks}
\\
Besides the one we showed in Fig.\ref{f2}{\bf b}, 
there are several ways to construct the 1D antiferromagnet with uniform DM interaction in a magnetic field. 
For completeness, we now provide a classification that applies to all of them. 
Let us start by reminding of a simple 1D antiferromagnet with only a Heisenberg exchange $J$. 
In contrast to the ferromagnets, it hosts two-fold degenerate magnon branches that cross at $k=0$ (see the top panel of Fig~\ref{f3}), 
and by the easy axis anisotropy, $\Lambda <0$, the gap opens, but the energy bands remain degenerate. 
The DM interaction perpendicular to the magnetic moments does not change the band structure, 
which is shown in Fig.~\ref{f3}{\bf a}. 
The $SU(2)$ symmetry is preserved for all these cases. 
By the introduction of $\bm D$ parallel to the magnetic moments, the symmetry reduces to $U^z(1)$. 
The two branches of bands shift in opposite directions \cite{okuma2017,cheng2016-2,gitgeatpong2017} as shown in Fig.~\ref{f3}{\bf b}.
These two cases are already studied in experiments\cite{gitgeatpong2017}. 
\par
We now examine the easy plane antiferromagnet in Fig.~\ref{f3}{\bf c}. 
When $\Lambda > 0$, one of the modes becomes gapped, 
which is responsible for the in-plane stretching mode. 
The gap of the remaining mode opens when the in-plane rotational symmetry is broken by $\bm h$.
Again, when we set $\bm D \perp \bm M_A, \bm M_B$ (upper panel), the magnon band structures remain unchanged. 
The direction of the spin carried by magnon does not vary with $\bm k$ since the system has 
$U(1)$ symmetry about the $x$-axis, which we denote $U^{x}(1)$. 
If we rotate $\bm M_A$ and $\bm M_B$ within the easy plane off the direction parallel to $\bm D$, 
the energy band is slightly modified by $\bm D$. 
These cases become important when we apply a field. 
\par
\par
By the application of $\bm h$, 
$\bm M_A$ and $\bm M_B$ are canted off the collinear alignment and 
gain a net moment in the field direction. 
The excitation against this weak ferromagnetic element couples to $\bm D$ depending on the 
relative angle between $\bm h$ and $\bm D$, 
which falls onto either of II$-$IV in Fig.~\ref{f3}{\bf d}. 
Case I is realized by applying a magnetic field parallel to the collinear magnetic moments 
with easy axis anisotropy. 
Its dispersion is actually observed experimentally in the noncentrosymmetric antiferromagnet, 
$\alpha$-Cu$_2$V$_2$O$_7$\cite{gitgeatpong2017}. 
In a finite magnetic field and for a noncollinear spin configuration, 
magnons start to carry spin moment $S_{\bm{k},\pm}$ that has a net value opposite to 
$\bm h \propto \bm M_A +\bm M_B$;  
\ch{Since $S_{\bm{k},\pm}$ 
is a linear combination of vectors $\bm M_A$ and $\bm M_B$ (Eq. (\ref{Smag})\cite{okuma2017}), 
having a noncollinear $\bm M_A$ and $\bm M_B$ is the {\it necessary} 
condition to vary both the direction and the amplitude of $\bm{S}_{\bm{k},\pm}$. }
Although this condition is fulfilled for Cases II-IV in Fig.\ref{f3}{\bf d}, 
only Case II breaks the $SU(2)$ symmetry down to $\{e\}$ and exhibits 
directional variation of $\bm S_{k,\pm}$ on $\bm k$. 
\par
In the other three cases, the spin texture is restricted by the symmetry of the BdG Hamiltonian; 
In Case I, the $U^{z}(1)$ symmetry leads to the quantization of $\bm{S}_{\bm{k},\pm}$. 
In Case II, the $U^{x}(1)$ symmetry allows $|\bm{S}_{\bm{k},\pm}|$ to vary, 
while they all point in the same direction following Eq.(\ref{eq:uni-S}). 
Case IV does not have a $U^{x}(1)$ symmetry, 
but $S_{\pm}(\bm k)$ again follows Eq.(\ref{eq:uni-S}). 
This is because the bosonic BdG Hamiltonian satisfies a similar condition, 
$[H_{\mathrm{SW}}(\bm{k}),J^{x}K]=0$, 
assisted by a complex conjugate operator, $K$. 
We denote this as $\bar{U}^{x}(1)$ symmetry for convenience. 
\\
Similar discussion can be developed for the ferromagnets with uniform DM interaction 
and for the antiferromagnets with staggered DM interaction, that completes the 
classification of the role of DM interactions on ferro and antiferromagnets 
and adds to Fig.\ref{f3}{\bf a}-{\bf c} the relationships with other cases (see Supplementary material C). 
For example, the antiferromagnet shown in Fig.\ref{f3}{\bf b} is then regarded as the combination of 
two ferromagnets with uniform DM interaction, each showing nonreciprocity in opposite directions 
which is explicitly shown in Supplementary Fig.S1. 
However, the magnetic moments are quantized on each of the magnon branches. 
\vspace{5mm}
\\
{\bf \small Designing 2D spin textures}
\\
Based on the above mentioned classification I$-$IV, one can design a spin texture of a 2D antiferromagnet by hand. 
Our Rashba and Dresselhaus type magnon spin textures offer a good example; 
Let us configure the direction, $\bm \delta_{\rm II}$, 
that reproduce case II in Fig.\ref{f3}{\bf d} along which the spin textures emerge in a most significant manner. 
This is done by taking the linear combination of $\bm \delta_1$ and $\bm \delta_2$, 
so as to have the combination of the two $\bm D$'s attached to them 
become perpendicular to $\bm h$ (see the black arrow in the middle panel of Fig.\ref{f1}{\bf b}). 
As $\bm h$ rotates clockwise, the direction of $\bm \delta_{\rm II}$ rotates anti-clockwise/clockwise for Rashba/Dresselhaus-type antiferromagnets. 
The other direction, $\bm \delta_{\rm III}$, that the locking does not occur is defined perpendicular to $\bm \delta_{\rm II}$, 
in a way that the combination of their two $\bm D_\parallel$'s points toward $\bm h$. 
We also put a constraint that the net moment points in the direction opposite to $\bm h$. 
These considerations allow us to figure out the overall textures without detailed calculation. 
\par
As one can anticipate from the evolution of spin texture, 
the band profile rotates following the field-angle $\phi=0\rightarrow 2\pi$, 
clockwise and anti-clockwise for Dresselhaus and Rashba-types, respectively. 
Accordingly, the nonreciprocity appears in the $(k_x,k_y)=(\cos\phi,-\sin\phi)$-direction for the 
Dresselhaus-type, and in the direction perpendicular to the field for the Rashba-type 
(see Supplementary Fig.S2). 
\vspace{5mm}
\\
\noindent
{\bf Discussion}
\\
{\bf \small Summary and outlook}\\
Based on the symmetry arguments and the idea of constructing the 2D antiferromagnets 
using the building blocks of 1D antiferromagnetic chains, a strategy to design a variety of spin textures is thus provided. 
Previously, active discussions on the physics of magnons were given in ferro or ferrimagnets with 
its magnetic moments {\it parallel} to the DM interactions; 
for the {\it staggered} DM interaction that does not break SIS, a topological magnon contributing to the thermal Hall effect is observed, 
and for uniform DM interaction, a nonreciprocity of ferromagnetic magnons were reported. 
Here, we clarified another aspect of magnons that the {\it uniform} DM interactions breaking global SIS in the antiferromagnet 
can generate as rich spin-textures as the Rashba and Dresselhaus electronic semiconductors. 
The two degenerate energy branches of the uniform antiferromagnet 
carry the magnetization pointing in the nearly opposite directions, 
which are regarded as the internal pseudo-spin degrees of freedom relevant to magnetic sublattices. 
The uniform DM interaction serves as a pseudo-spin orbit coupling of magnons 
and generate a spin texture over the whole reciprocal space. 
To activate such fictitious pseudo-SO coupling of magnons, 
an interplay with a magnetic field and a spin anisotropy plays a crucial role, 
in order to fully break the pseudo-spin-$SU(2)$ symmetry. 
The resultant textures are easily controlled by the magnetic field angle. 
\par
The 2D Dresselhaus antiferromagnet is actually realized in a noncentrosymmetric 
spin-$5/2$ antiferromagnet Ba$_2$MnGe$_2$O$_7$, with space group $P\bar{4}2_{1}m$.
It undergoes a N\'eel transition at $T_N=4$K into an easy plane type antiferromagnetic phase\cite{masuda2010,murakawa2012}, 
where the microwave non-reciprocity is indeed observed\cite{iguchi2018}. 
A spin-3/2 multiferroic Ba$_2$CoGe$_2$O$_7$\cite{murakawa2010,yi2008}, possibly of space group P$\bar{4}$2$_1 m$, 
is considered to have a similar property. 
The exchange interactions is Ba$_2$MnGe$_2$O$_7$ is $J\sim$ 27$\mu$eV\cite{masuda2010}, 
which is much smaller than the other materials of the same family, 
possibly allowing for the examination of the present phenomena by several experimental probes, 
and with a very small magnetic field of less than few tesla. 
\ch{Our theory shows that even for the materials with much larger $J$, one can obtain 
a similar spin texture by setting $h < 0.1J$ (see Fig.\ref{f4}{\bf c} and Supplementary Fig.S3)}. 
Besides such noncentrosymmetric antiferromagnets, 
the interface of typical centrosymmetric antiferromagnets is expected to show magnonic spin momentum locking. 
Thus, the scheme we proposed may allow a strong command of designing spin textures 
toward the application for antiferromagnetic spintronics\cite{junwirth2018}. 
\par
Although spin textures and nonreciprocity were difficult to realize 
within previous considerations on antiferromagnets, 
there is one rare example on a honeycomb lattice\cite{cheng2016,zyuzin2016} exhibiting nonreciprocity\cite{hayami2016} 
with toroidal ordering, and thermal Hall effect\cite{owerre2017,owerre2017-2}. 
The crystal structure of the material keeps the SIS, and the DM interaction acts on next nearest neighboring 
sites belonging to the same sublattice, and align in the staggered manner. 
In our scheme, it can be regarded as a staggered assembly of 1D ferromagnets with uniform DM interactions, 
that may provide a generalized interpretation on its properties (see Supplementary Fig.S1{\bf g}). 
\vspace{3mm}
\\
\noindent
{\bf \small Device application}
\\
The advantage of having spin texture in insulators is to store directly information by controlling the texture itself, 
and how to characterize and read the information is the important and challenging part of device applications. 
On the top of that, it is starting to be recognized that antiferromagnets have several advantages 
for a memory storage device over ferromagnets\cite{junwirth2018}. 
Since the antiferromagnetic state does not generate stray fringing fields, it is robust against any perturbation. 
Writing speed of memory is generally limited by the resonance frequency, 
which can be much higher in the antiferromagnets than in ferromagnets. 
Nevertheless, the potential abilities of antiferromagnets remained unexplored. 
\par
Our results in Fig.\ref{f1}{\bf b} indicate that the magnonic spin current can be a good probe for spin textures: 
the magnetic moment $\bm S_{\bm k,\pm}$ at relatively large $\bm k$ depends on the sign of $\bm k$. 
Therefore, by the experimental setup shown in Fig.~\ref{f4}{\bf a}, 
one can selectively excite magnons that generate a nearly pure spin current. 
The electromagnetic waves of frequency $f\sim 10^1-10^3$ GHz emitted from a micro antenna 
on the antiferromagnetic sample generates antiferromagnetic magnons with wave vectors $+\bm k$ and $-\bm k$, 
which have an amplitude typically of an inverse of antenna width $d$. 
\ch{The group velocity of the excited magnons, $d\omega_{\bm k,-}/dk$, together with 
the amount of excited spin density $|\bm S_{\bm k,-}\!-\!\bm S_{-\bm k,-}|$ 
is shown as functions of $k$ in Figs.\ref{f4}{\bf b} and \ref{f4}{\bf c}.
The gradual increase of spin current with $|k|$ and $D$ is common to noncentrosymmetric ferromagnets\cite{iguchi15}. 
The group velocity shows a sharp and linear increase from $k=0$ and saturates at small $k$, 
which is the feature of the antiferromagnetic magnon with nearly linear dispersion at small $k$, 
and it works as an advantage for the experimental observation.   
} 
By probing the direction of spin current through an inverse spin Hall voltage of Pt electrode, 
one can figure out the structure of the texture in a rotating field. 
\ch{Unlike the diffusive spin current carried by conduction electrons in the presence of SO coupling, 
the magnon spin current in insulators are stable regardless of the $D$ and $\Lambda$-terms\cite{Ruckriegel2018}, 
as far as the direct couplings with lattices or impurities are not included. 
For the dissipation due to such an extrinsic effect, 
the phenomenological theories will give the evaluation of the propagation length, 
which is sensitive to circumstances\cite{Ruckriegel2017}. 
}
The present study provides a roadmap to electrically detect and control spin textures in ubiquitous antiferromagnets, 
opening up a pathway for antiferromagnetic spintronics. 
\vspace{5mm}
\\
\noindent
{\bf Methods}
\\
We perform a spin wave analysis on two sublattice antiferromagnets. 
The starting point is a collinear or a canted antiferromagnetic classical order. 
We apply a local unitary transformation, $U^\dagger{\cal H} U$, and set the $z$-axis of the spin operator space to 
the direction of the magnetic moments.
For example, the unitary operator for 2D easy plane antiferromagnets on a square lattice shown 
in Fig.\ref{f1}{\bf a} is given as 
\begin{equation}
U=\bigotimes_{j}\mathrm{exp}\left(i\frac{\pi}{2}S_{j}^{x}\right)
\mathrm{exp}\left(-i(\theta_M\mathrm{e}^{i\bm{Q}\cdot\bm{r}_{j}}-\phi)S_{j}^{y}\right), 
\label{eq:unitary}
\end{equation}
which sets the spin operators in fictitious local space antiparallel. 
Here, $\bm{r}_{j}$, is the spatial coordinate of site-$j$, 
and the ordering wave vector $\bm{Q}$ satisfies ${\rm e}^{i\bm Q\bm r_j}=\pm 1$ for sublattices A and B, respectively. 
The linearized Holstein-Primakoff transformation is given as 
\begin{equation}
S_{i}^{+}\simeq\sqrt{2S}a_{i}
\hspace{10pt}
S_{i}^{-}\simeq\sqrt{2S}a_{i}^{\dagger}
\hspace{10pt}
S_{i}^{z}=S-a_{i}^{\dagger}a_{i},
\end{equation}
for sublattice-A, and
\begin{equation}
S_{j}^{+}\simeq\sqrt{2S}b_{j}^{\dagger}
\hspace{10pt}
S_{j}^{-}\simeq\sqrt{2S}b_{j}
\hspace{10pt}
S_{j}^{z}=-S+b_{j}^{\dagger}b_{j}
\end{equation}
for sublattice-B. 
After the Holstein-Primakoff and Fourier transformations, 
$U^{\dagger}\mathcal{H}U\simeq\mathrm{const}+\mathcal{H}_{\mathrm{SW}}$, 
the spin wave Hamiltonian is written finally in the quadratic form as Eq.(\ref{eq:uhu}).
The bosonic BdG Hamiltonian is written as 
\begin{equation}
H_{\mathrm{BdG}}(\bm{k})=
\begin{pmatrix}
\Xi_{\bm{k}} & \Delta_{\bm{k}}\\
\Delta_{-\bm{k}}^{*} & \Xi_{-\bm{k}}^{*}
\end{pmatrix}.
\end{equation}
Here $\Xi_{\bm{k}}$ and $\Delta_{\bm{k}}$ are $2\times2$ matrices satisfying $\Xi_{\bm{k}}^{\dagger}=\Xi_{\bm{k}}$ and $\Delta_{-\bm{k}}^{\dagger}=\Delta_{\bm{k}}^{*}$. 
$H_{\mathrm{BdG}}(\bm{k})$ is diagonalized analytically using the paraunitary matrix \cite{colpa1978}. 
The actual diagonalization is done by solving the following eigenvalue equation, 
\begin{equation}
\Sigma^{z}H_{\mathrm{BdG}}(\bm{k})\bm{t}_{\pm}(\bm{k})=\omega_{\pm}(\bm{k})\bm{t}_{\pm}(\bm{k})
\label{eq:egeq}
\end{equation}
with $\Sigma^{z}=\tau^{z}\otimes\sigma^{0}$. 
The eigenvectors of magnons, $\bm{t}_{\pm}(\bm{k})=(u_{\pm,A}(\bm{k}), u_{\pm,B}(\bm{k}), v_{\pm,A}(\bm{k}), v_{\pm,B}(\bm{k}))^{T}$, 
satisfy the normalization condition, 
$\bm{t}_{\eta}^{\dagger}(\bm{k})\Sigma^{z}\bm{t}_{\eta'}(\bm{k})=\delta_{\eta,\eta'}$
with $\eta,\eta'=\pm$.
We obtain the analytical form of the magnon dispersions $\omega_{\pm}(\bm{k})$ and corresponding eigenvectors $\bm{t}_{\pm}(\bm{k})$ consisting of two branches.
The local spectral weight used in Eq.(\ref{Smag}) is given as 
\begin{equation}
d_{A/B}(\omega_\pm(\bm k))= |u_{\pm,A/B}(\bm k)|^2+|v_{\pm,A/B}(\bm k)|^2 . 
\end{equation}
\begin{acknowledgments}
We acknowledge discussions with Y. Iguchi and Y. Nii and K. Penc. 
This work is supported by JSPS KAKENHI Grant Numbers JP17K05533, JP17K05497, JP17H02916, 
JP16H04008 and JP17H05176. 
\end{acknowledgments}

\vspace{5mm}

\end{document}


\renewcommand{\figurename}{Figure S\!}

\title{Supplementary material for ``Designing Rashba-Dresselhaus effect in magnetic insulators"}

\maketitle
\noindent
\subsection{Symmetry-based analysis}
We showed in the main text that two-sublattice antiferromagnets following the Hamiltonian $\mathcal{H}$ in Eq.(1) 
is transformed to the BdG Hamiltonian as, 
$U^{\dagger}\mathcal{H}U\simeq\mathrm{const}+\mathcal{H}_{\mathrm{SW}}$, 
and that the spin wave Hamiltonian takes the form using 
$\Phi_{\bm{k}}=(a_{\bm{k}}, b_{\bm{k}}, a_{-\bm{k}}^{\dagger}, b_{-\bm{k}}^{\dagger})$ as, 
\begin{eqnarray}
&&\mathcal{H}_{\mathrm{SW}}=\frac{1}{2}\sum_{\bm{k}}\Phi_{\bm{k}}^{\dagger}H_{\mathrm{BdG}}(\bm{k})\Phi_{\bm{k}}, 
\nonumber\\
&& H_{\mathrm{BdG}}(\bm{k})=
\begin{pmatrix}
\Xi_{\bm{k}} & \Delta_{\bm{k}}\\
\Delta_{-\bm{k}}^{*} & \Xi_{-\bm{k}}^{*}
\end{pmatrix}.
\end{eqnarray}
The $2\times2$ matrices, $\Xi_{\bm{k}}$ and $\Delta_{\bm{k}}$, 
satisfy $\Xi_{\bm{k}}^{\dagger}=\Xi_{\bm{k}}$ and $\Delta_{-\bm{k}}^{\dagger}=\Delta_{\bm{k}}^{*}$, 
and can be written in the general form as, 
\begin{equation}
\Xi_{\bm{k}}=\sum_{\mu=0,x,y,z}\xi_{\bm{k}}^{(\mu)}\sigma^{\mu},
\hspace{20pt}
\Delta_{\bm{k}}=\sum_{\mu=0,x,y,z}\Delta_{\bm{k}}^{(\mu)}\sigma^{\mu},
\end{equation}
where $\xi_{\bm{k}}^{(\mu)}\in\mathbb{R}$, $\Delta_{\bm{k}}^{(\mu)}\in\mathbb{C}$, 
and $\sigma^{\mu}$ is the unit and Pauli matrix acts on a sublattice space ($\mu=0,x,y,z$).
In addition, $\Delta_{\bm{k}}^{(\mu)}$ satisfy the following relations, 
\begin{equation}
\Delta_{-\bm{k}}^{(0)}=\Delta_{\bm{k}}^{(0)},
\hspace{10pt}
\Delta_{-\bm{k}}^{(x)}=\Delta_{\bm{k}}^{(x)},
\hspace{10pt}
\Delta_{-\bm{k}}^{(y)}=-\Delta_{\bm{k}}^{(y)},
\hspace{10pt}
\Delta_{-\bm{k}}^{(z)}=\Delta_{\bm{k}}^{(z)}.
\end{equation}
In Eq.(5), we introduced the $SU(2)$ operator which acts on the product space of the particle-hole and sublattice space, 
which we rewrite here for convenience as, 
\begin{equation}
J^{x}=\tau^{0}\otimes\sigma^{x},
\hspace{20pt}
J^{y}=\tau^{z}\otimes\sigma^{y},
\hspace{20pt}
J^{z}=\tau^{z}\otimes\sigma^{z}. 
\nonumber
\end{equation}
In the present system, the reduction of $SU(2)$ symmetry to $U(1)$ about the $x$ (and $y$) and $z$-axes 
have different physical implication. 
Therefore, we denote the $U(1)$ symmetry about the $\mu$-axis as $U^\mu(1)$ for convenience. 
If the system has the $U^{x}(1)$ symmetry,
a following relation holds from $[H_{\mathrm{BdG}}(\bm{k}),J^{x}]=0$;
\begin{equation}
[\Xi_{\bm{k}},\sigma^{x}]=0,
\hspace{20pt}
[\Delta_{\bm{k}},\sigma^{x}]=0.
\label{Ux}
\end{equation}
Similarly, if the system has the $U^\mu(1)$ symmetry for each $\mu=y,z$, 
a following relation holds from $[H_{\mathrm{BdG}}(\bm{k}),J^{\mu}]=0$; 
\begin{equation}
[\Xi_{\bm{k}},\sigma^{\mu}]=0,
\hspace{20pt}
\{\Delta_{\bm{k}},\sigma^{\mu}\}=0.
\label{Umu}
\end{equation}
\par
Let us first consider $SU(2)$-symmetric cases, $[H_{\mathrm{BdG}}^{SU(2)}(\bm{k}),J^{\mu}]=0$ for all $\mu=x,y,z$.
In such cases,
$\Xi_{\bm{k}}$ and $\Delta_{\bm{k}}$ satisfy Eq.(\ref{Ux}) and (\ref{Umu}), 
leading to the following form of $\Xi_{\bm{k}}$ and $\Delta_{\bm{k}}$, 
\begin{equation}
\Xi_{\bm{k}}=\xi_{\bm{k}}^{(0)}\sigma^{0},
\hspace{20pt}
\Delta_{\bm{k}}=\Delta_{\bm{k}}^{(x)}\sigma^{x},
\end{equation}
Then, the magnon bands, $\omega_{\pm}(\bm{k})$, become doubly degenerate and take the following form,
\begin{equation}
\omega_{\pm}(\bm{k})=\omega(\bm{k})=\frac{\xi_{\bm{k}}^{(0)}-\xi_{-\bm{k}}^{(0)}}{2}+\sqrt{\left(\frac{\xi_{\bm{k}}^{(0)}+\xi_{-\bm{k}}^{(0)}}{2}\right)^{2}-\left|\Delta_{\bm{k}}^{(x)}\right|^{2}}.
\label{omega-su2}
\end{equation}
An example of the $SU(2)$ symmetric system is shown in Fig .1{\bf b} with $D=h=0$,
whose bosonic BdG Hamiltonian is written as
\begin{equation}
H_{\mathrm{BdG}}^{SU(2)}(k)=(\tau^{0}\otimes\sigma^{0})2JS+(\tau^{x}\otimes\sigma^{x})2JS\mathrm{cos}k.
\end{equation}
One can easily check that this Hamiltonian satisfies $[H_{\mathrm{BdG}}(k),J^{\mu}]=0$ for $\mu=x,y,z$,
and the magnon bands are given by Eq.(\ref{omega-su2}) with $\xi_{k}^{(0)}=2JS$ and $\Delta_{k}^{(x)}=2JS\mathrm{cos}k$.
\par
Next, we consider $U^{x}(1)$ symmetric cases. 
The magnon bands are generally not degenerate, and we can choose the eigenvectors of magnons, 
$\bm{t}_{\pm}(\bm{k})$, as the eigenvectors of $J^{x}$
since this operator commutes with $\Sigma^{z}$,
\begin{equation}
J^{x}\bm{t}_{\pm}(\bm{k})=\lambda_{\pm}\bm{t}_{\pm}(\bm{k}).
\label{eq-Jx}
\end{equation}
From this equation, we find 
\begin{equation}
\bm{t}_{\pm}^{\dagger}(\bm{k})(J^{x})^{\dagger}\Sigma^{z}=\lambda_{\pm}^{*}\bm{t}_{\pm}^{\dagger}(\bm{k})\Sigma^{z}.
\label{eq-Jx-dag}
\end{equation}
The combination of Eq. (\ref{eq-Jx}) and (\ref{eq-Jx-dag}) leads to $|\lambda_{\pm}|=1$. 
Thus, a following relation holds from Eq.(\ref{eq-Jx}),
\begin{equation}
|u_{\pm,\mathrm{A}}(\bm{k})|=|u_{\pm,\mathrm{B}}(\bm{k})|,
\hspace{20pt}
|v_{\pm,\mathrm{A}}(\bm{k})|=|v_{\pm,\mathrm{B}}(\bm{k})|.
\label{A=B}
\end{equation}
From Eq.(\ref{A=B}), one can immediately see $d_{\mathrm{A}}(\omega_{\pm}(\bm{k}))=d_{\mathrm{B}}(\omega_{\pm}(\bm{k}))=d(\omega_{\pm}(\bm{k}))$,
and spins carried by magnons, $\bm{S}_{\bm{k},\pm}$, take the following form,
\begin{equation}
\bm{S}_{\bm{k},\pm}=-(\bm{M}_{\mathrm{A}}+\bm{M}_{\mathrm{B}})d(\omega_{\pm}(\bm{k})).
\label{uniS}
\end{equation}
Only the amplitude of $\bm{S}_{\bm{k},\pm}$ depends on $\bm{k}$, 
and the direction of $\bm{S}_{\bm{k},\pm}$ is always opposite to the sum, 
$\bm{M}_{\mathrm{A}}+\bm{M}_{\mathrm{B}}$. 
\par
An example of the $U^{x}(1)$ symmetric system is the 2D antiferromagnet with $\bm{D}=\bm{0}$, 
and the parameters of the BdG Hamiltonian is given as, 
\begin{eqnarray}
\xi_{\bm{k}}^{(0)}&=&4JS\mathrm{cos}2\theta_M+\Lambda S+h\mathrm{sin}\theta_M,\\
\xi_{\bm{k}}^{(x)}&=&-4JS\gamma_{\bm{k}}\mathrm{sin}^{2}\theta_M-2DSg_{1,\bm{k}}\mathrm{sin}\theta_M,\\
\Delta_{\bm{k}}^{(0)}&=&-\Lambda S,\\
\Delta_{\bm{k}}^{(x)}&=&4JS\gamma_{\bm{k}}\mathrm{cos}^{2}\theta_M,\\
\Delta_{\bm{k}}^{(y)}&=&-i2DSg_{2,\bm{k}}\mathrm{cos}\theta_M,
\end{eqnarray}
\begin{equation}
\gamma_{\bm{k}}=\frac{\mathrm{cos}\bm{k}\cdot\bm{\delta}_{1}+\mathrm{cos}\bm{k}\cdot\bm{\delta}_{2}}{2},\\
\end{equation}
\begin{eqnarray}
g_{1,\bm{k}}&=&\mathrm{cos}\phi\frac{\mathrm{sin}\bm{k}\cdot\bm{\delta}_{1}+\mathrm{sin}\bm{k}\cdot\bm{\delta}_{2}}{\sqrt{2}}-\mathrm{sin}\phi\frac{\mathrm{sin}\bm{k}\cdot\bm{\delta}_{1}-\mathrm{sin}\bm{k}\cdot\bm{\delta}_{2}}{\sqrt{2}},
\nonumber\\
\\
g_{2,\bm{k}}&=&\mathrm{cos}\phi\frac{\mathrm{sin}\bm{k}\cdot\bm{\delta}_{1}-\mathrm{sin}\bm{k}\cdot\bm{\delta}_{2}}{\sqrt{2}}+\mathrm{sin}\phi\frac{\mathrm{sin}\bm{k}\cdot\bm{\delta}_{1}+\mathrm{sin}\bm{k}\cdot\bm{\delta}_{2}}{\sqrt{2}}.
\nonumber\\
\end{eqnarray}
The canting angle is given as $\mathrm{sin}\theta_M=h/8JS$.
One can easily check that the bosonic BdG Hamiltonian with $D=0$ has $U^{x}(1)$ symmetry,
and the finite $D$ reduces the $U^{x}(1)$ symmetry down to $\{e\}$.
Note that cases with $U^{y}(1)$ symmetry are basically the same and Eq.(\ref{uniS}) holds.
\par
Finally, we consider cases with $U^{z}(1)$-symmetry. 
$\Xi_{\bm{k}}$ and $\Delta_{\bm{k}}$ satisfy Eq.(\ref{Umu}) with $\mu=z$ 
and take the following form, 
\begin{equation}
\Xi_{\bm{k}}=\xi_{\bm{k}}^{(0)}\sigma^{0}+\xi_{\bm{k}}^{(z)}\sigma^{z},
\hspace{20pt}
\Delta_{\bm{k}}=\Delta_{\bm{k}}^{(x)}\sigma^{x}+\Delta_{\bm{k}}^{(y)}\sigma^{y}
\label{xidelta}
\end{equation}
The magnon bands are written as, 
\begin{eqnarray}
&&\omega_{\pm}(\bm{k})
=\frac{\xi_{\bm{k}}^{(0)}-\xi_{-\bm{k}}^{(0)}}{2}\pm\frac{\xi_{\bm{k}}^{(z)}
   +\xi_{-\bm{k}}^{(z)}}{2}
\rule{30mm}{0mm}
\nonumber \\
&& \hspace{2mm}+\sqrt{\left(\frac{\xi_{\bm{k}}^{(0)}\!+\!\xi_{-\bm{k}}^{(0)}}{2}
     \pm\frac{\xi_{\bm{k}}^{(z)}\!-\!\xi_{-\bm{k}}^{(z)}}{2}\right)^{2}\hspace{-2pt}-\hspace{2pt}
     \!\-\!\left|\Delta_{\bm{k}}^{(x)}\!\mp\!i\Delta_{\bm{k}}^{(y)}\right|^{2}}
\end{eqnarray}
One can see that $\xi_{\bm{k}}^{(z)}$ and $\Delta_{\bm{k}}^{(y)}$ break the $SU(2)$ symmetry and split the degenerate magnon bands.
The corresponding eigenvectors are also the eigenvectors of $J^{z}$,
and the explicit form of these eigenvectors are
\begin{equation}
\hspace*{-3mm}
\bm{t}_{+}(\bm{k})\!=\!
\begin{pmatrix}
\mathrm{cosh}\chi_{+}(\bm{k})\\
0\\
0\\
\mathrm{e}^{i\rho_{+}(\bm{k})}\mathrm{sinh}\chi_{+}(\bm{k})
\end{pmatrix},
\hspace{1pt}
\bm{t}_{-}(\bm{k})\!=\!
\begin{pmatrix}
0\\
\mathrm{cosh}\chi_{-}(\bm{k})\\
\mathrm{e}^{i\rho_{-}(\bm{k})}\mathrm{sinh}\chi_{-}(\bm{k})\\
0
\end{pmatrix}\!,
\rule{0mm}{2mm}
\end{equation}
where
\begin{eqnarray}
&&\mathrm{tanh}\chi_{\pm}(\bm{k})
=\frac{\omega_{\pm}(\bm{k})-\xi_{\bm{k}}^{(0)}\mp\xi_{\bm{k}}^{(z)}}{|\Delta_{\bm{k}}^{(x)}\mp i\Delta_{\bm{k}}^{(y)}|},
\nonumber \\
&&\mathrm{e}^{-i\rho_{\pm}(\bm{k})}
=\frac{\Delta_{\bm{k}}^{(x)}\mp i\Delta_{\bm{k}}^{(y)}}{|\Delta_{\bm{k}}^{(x)}\mp i\Delta_{\bm{k}}^{(y)}|}.
\end{eqnarray}
In such cases,
spins carried by magnons, $\bm{S}_{\bm{k},\pm}$, are
\begin{equation}
\bm{S}_{\bm{k},\pm}=\mp\frac{\bm{M}_{\mathrm{A}}-\bm{M}_{\mathrm{B}}}{2}-\frac{\bm{M}_{\mathrm{A}}+\bm{M}_{\mathrm{B}}}{2}\mathrm{cosh}2\chi_{\pm}(\bm{k}).
\end{equation}
This equation implies that the direction of spins varies with $\bm{k}$ 
if the classical spins in the ground state are non-collinear, $\bm{M}_{\mathrm{A}}+\bm{M}_{\mathrm{B}}\neq\bm{0}$.
However, $\bm{M}_{\mathrm{A}}$ and $\bm{M}_{\mathrm{B}}$ are collinear when the system has $U^{z}(1)$ symmetry. 
Therefore, $\bm{S}_{\bm{k},\pm}$ is quantized to, 
\begin{equation}
\bm{S}_{\bm{k},\pm}=\mp \bm{M}_{\mathrm{A}}= \pm \bm{M}_{\mathrm{B}}. 
\label{quantized-S}
\end{equation}
An example of the $U^{z}(1)$ symmetric system is shown in Fig.1{\bf b} 
with $D\neq0$ and $h=0$, 
whose bosonic BdG Hamiltonian is given by 
\begin{eqnarray}
H_{\mathrm{BdG}}(k)&=&(\tau^{0}\otimes\sigma^{0})2(J+\Lambda)S +(\tau^{x}\otimes\sigma^{x})2JS\mathrm{cos}k
 \nonumber \\
&& -(\tau^{y}\otimes\sigma^{y})2DS\mathrm{sin}k, 
\end{eqnarray}
which is connected to Eq.(\ref{xidelta}) via 
$\xi_{k}^{(0)}=2(J+\Lambda)S$, $\xi_{k}^{(z)}=0$, $\Delta_{k}^{(x)}=2JS\mathrm{cos}k$, 
and $\Delta_{k}^{(y)}=i2DS\mathrm{sin}k$. 
\par
In an applied field, $h\neq0$,
the bosonic BdG Hamiltonian is given as
\begin{eqnarray}
H_{\mathrm{BdG}}(k)&=&(\tau^{0}\otimes\sigma^{0})\xi_{k}^{(0)}+(\tau^{0}\otimes\sigma^{x})\xi_{k}^{(x)}
\nonumber \\
&& +(\tau^{x}\otimes\sigma^{0})\Delta_{k}^{(0)}+(\tau^{x}\otimes\sigma^{x})\Delta_{k}^{(x)}
\nonumber \\
&& +(\tau^{y}\otimes\sigma^{y})\Delta_{k}^{(y)},
\end{eqnarray}
where
\begin{eqnarray}
\xi_{k}^{(0)}&=&2JS\mathrm{cos}2\theta_M+\Lambda S(2-3\mathrm{sin}^{2}\theta_M)+h\mathrm{sin}\theta_M,
\rule{10mm}{0mm}
\\
\xi_{k}^{(x)}&=&-2JS\mathrm{sin}^{2}\theta_M\mathrm{cos}k,\\
\Delta_{k}^{(0)}&=&\Lambda S\mathrm{sin}^{2}\theta_M,\\
\Delta_{k}^{(x)}&=&2JS\mathrm{cos}^{2}\theta_M\mathrm{cos}k,\\
\Delta_{k}^{(y)}&=&i2DS\mathrm{cos}\theta_M\mathrm{sin}k,
\end{eqnarray}
and $\mathrm{sin}\theta_M=h/(4JS+2\Lambda S)$ is a canting angle.
This bosonic BdG Hamiltonian does not have $U(1)$ symmetry,
and the amplitude and the direction of the spin moments, $\bm{S}_{\bm{k},\pm}$, vary with $\bm k$ over the Brillouin zone.
\par
\subsection{Condition for spin textures}
We discuss the relevance of our work with 
Ref.[35] [N. Okuma, Phys. Rev. Lett. {\bf 119}, 107205 (2017)] 
and clarify the difference between the two. 
As we discussed in the main text, 
the spin momentum locking generally refers to the fixing of angle between 
particle momentum $\bm k$ and spin momentum $\bm S_{\bm k,\pm}$, but does not necessarily lead to 
the emergent spin texture itself, 
and is classified into three types, 
(i) $\bm{S}_{\bm{k},\pm}$ is quantized,
(ii) amplitude of $\bm{S}_{\bm{k},\pm}$ depends on $\bm{k}$ but the direction does not change, 
(iii) the amplitude and the direction of $\bm{S}_{\bm{k},\pm}$ vary with $\bm{k}$.
\par
Okuma assumed the spin Hamiltonian with rotational symmetry and considered 
the ground state with long range order, partially breaking the symmetry of the Hamiltonian. 
He shows that if there is an unbroken generator $S_{\mathrm{tot}}^{\mu}$, 
which fulfills $[\mathcal{H},S_{\mathrm{tot}}^{\mu}]=0$ after spontaneous symmetry breaking, 
it leads to the quantization of $\bm{S}_{\bm{k},\pm}$ as in Eq.(\ref{quantized-S}). 
\par
The existence of unbroken generator distinguishes (i) from (ii)/(iii), and provides the sufficient condition to have (i). 
However, it does not give any information to realize (ii) and (iii). 
Our symmetry-based discussion can also distinguish (ii) and (iii), 
and can detect the interesting spin textures, (iii), 
just by examining the degree of breaking of the $SU(2)$ symmetry of the general bosonic BdG Hamiltonian. 
A more intuitive explanation to distinguish (iii) from (i) and (ii) is 
to have different magnitude of component of $\bm{M}_{\mathrm{A}}$ and $\bm{M}_{\mathrm{B}}$ parallel to $\bm D$, 
which is nothing but Case II in Fig.3{\bf d} in the main text. 
\par
We note here briefly that the kagome lattice model discussed by Okuma assumes the ground state with 120 degree N\'eel ordering. 
While he showed that the magnons defined on a classically ordered moments on three sublattices 
contribute to the emergent spin texture, such three sublattice magnetic structure is unstable agaisnt 
gapless excitations known as ``weathervane mode" or ``line defects"\cite{chalker}, and is hardly realized. 
(They are the excitations of spins particular to the frustrated kagome lattice that has a texture of weathervane or line). 
Even this kind of ground state were able to be stabilized by some artificial parameters, 
the final state realized by the symmetry reduction from $U^z(1)$ to $\{ e\}$ is beyond his framework. 
Our analysis can be extended to systems with more than two magnetic sublattices in the unit cell, 
which could then be applied to that kind of case as well. 

\par
\begin{figure*}[tbp]
\includegraphics[width=17cm]{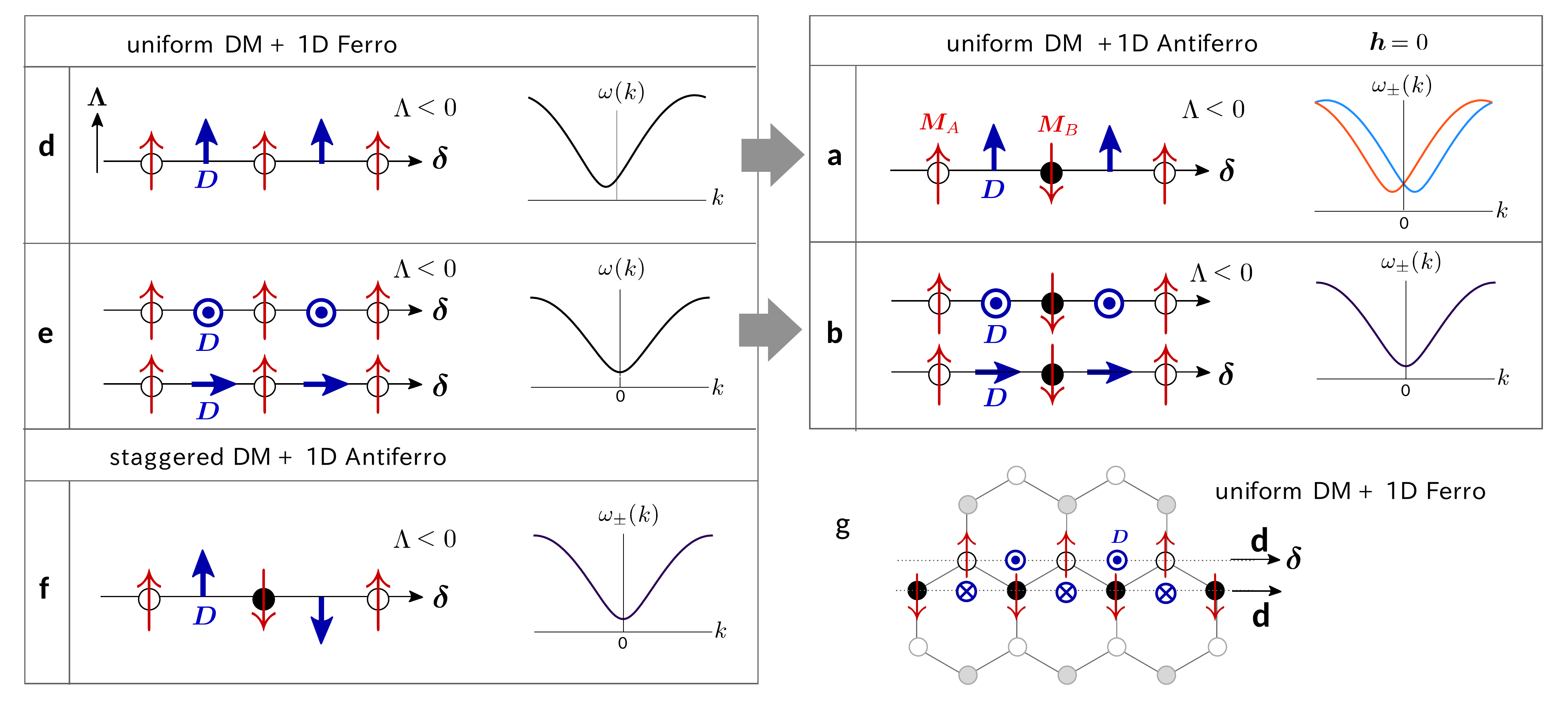}
\caption{({\bf d})-({\bf f}) Schematic illustration of the 1D magnetic chain with easy-axis anisotropy ($\Lambda<0$); 
({\bf d,e}) Ferromagnetic chain with uniform DM interaction along three different directions, 
and ({\bf f}) antiferromagnetic chain with staggered DM interactions. 
Panels ({\bf a},{\bf b}) are the antiferromagnetic chain with uniform DM interactions, 
which is the same as Figs.3{\bf a} and {\bf b} in the main text. 
Magnon bands calculated for $\bm h=0$ and $\bm D\ne 0$ is shown, where only 
the ferromagnetic case ({\bf d}) shows nonreciprocity. 
({\bf g}) Antiferromagnetic honeycomb lattice, which is understood as a combination of ferromagnetic chains with uniform DM 
with magnetic moments belonging to different sublattices pointing in the opposite direction. 
}
\label{fS1}
\end{figure*}
\begin{figure*}[tbp]
\includegraphics[width=18cm]{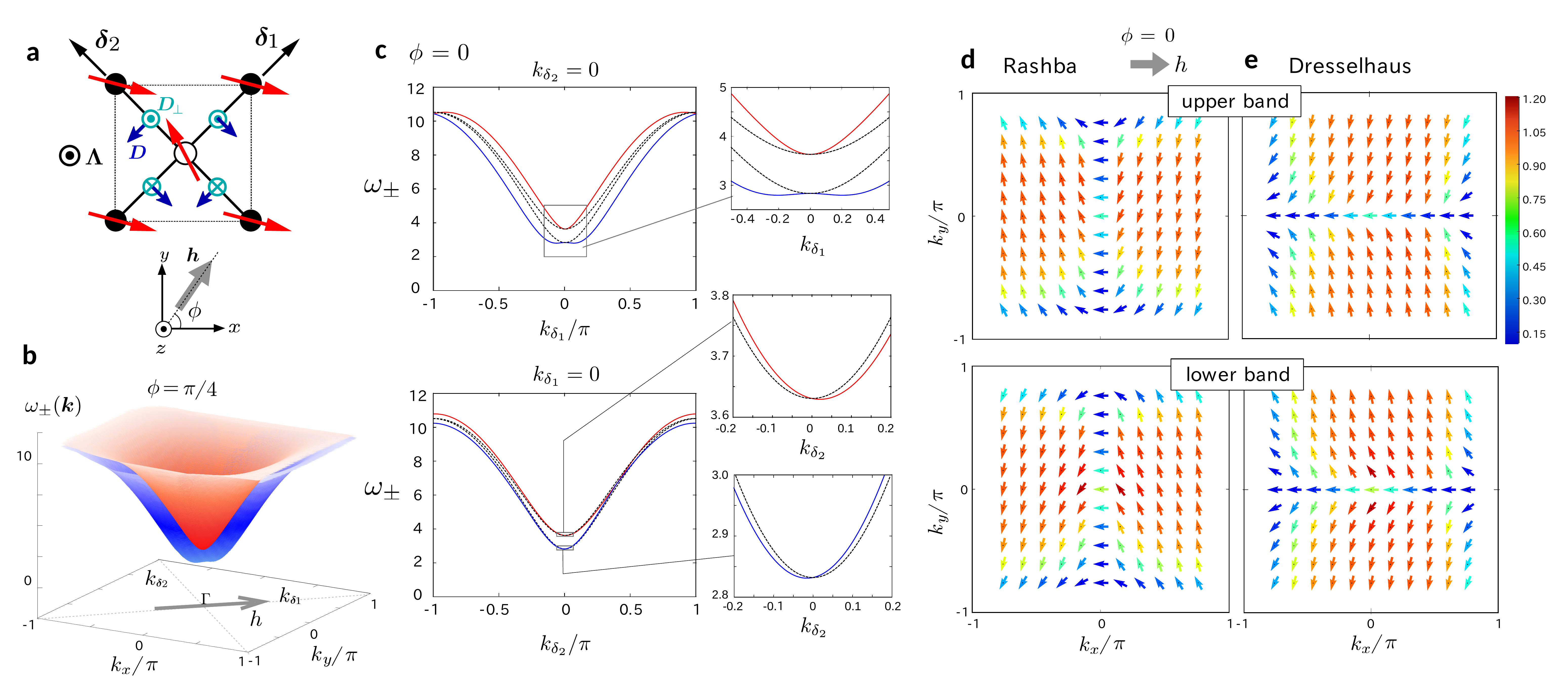}
\caption{({\bf a}) 2D SIS broken Dresselhaus antiferromagnet with easy-$xy$-plane anisotropy $\Lambda>0$. 
It is the same as we showed in Fig.1{\bf a} in the main text besides the staggered DM interaction normal to the plane, $D_\perp$. 
({\bf b}) Magnon band dispersion of the model in panel {\bf a} 
with $J=1$ $D_{\perp}=D=0.2$, $\Lambda=0.05$, 
and $h=2.0$ at $\phi=\pi/4$. 
The magnon bands have a profile that breaks the rotational symmetry, and rotates following the field angle $\phi$. 
({\bf c}) Cross sections of the bands in panel {\bf a}
along $\bm k_{\bm \delta_1} (\parallel \bm h)$ ($k_{\bm \delta_2}=0$) and 
$k_{\bm \delta_2} (\perp \bm h)$ ($k_{\bm \delta_1}=0$) lines. 
Solid and broken lines denote the magnon bands at $D\ne 0$ and $D=0$, respectively. 
The dispersions around the $\Gamma$ point is magnified in the right panels. 
When $D\neq 0$ the magnon bands become nonreciprocal 
in the direction perpendicular to $\bm h$ corresponding to case III in Fig.3{\bf d}. 
The $\bm h$-direction corresponds to case II. 
({\bf d,e}) Spin textures $\bm S_{\bm k,\pm}$ of Rashba and Dresselhaus magnons with magnetic field pointing in the $x$-direction ($\phi=0$). The upper band textures are shown to be compared with the lower band ones already discussed in the main text. 
}
\label{fS2}
\end{figure*}
\subsection{1D ferromagnets and antiferromagnets}
In the main text (Fig.3(a)-(c)), we dealt with the 1D antiferromagnets with uniform DM interactions. 
Here, we consider other 1D cases described by the Hamiltonian Eq.(1) in the main text: 
ferromagnets and antiferromagnets with staggered DM interactions. 
\par
Let us start from a simplest uniform ferromagnet with a gapless single cosine dispersion which is parabolic at $\bm k=0$, 
and by the easy axis anisotropy, $\Lambda <0$, the gap opens. 
Figures~S\ref{fS1}{\bf d} and S\ref{fS1}{\bf e} shows three different ways of introducing uniform DM interaction, 
and among them, only the vector $\bm D$ parallel to the ordered magnetic moments couples 
to the orbital motion of magnons and contributes to the nonreciprocity\cite{melcher1973,kataoka1987}, 
i.e., the bands are shifted off the center as, 
$\omega(\bm k) \ne \omega(-\bm k)$. 
There, the off-diagonal ($xy$) element of Eq.(1) is transformed in the language of bosons as, 
$\big(J\bm S_i \cdot \bm S_j +\bm D_{ij} \cdot (\bm S_i \times \bm S_j)\big)_{xy}
\propto ({\rm e}^{-i \varphi}a_i^\dagger a_j + {\rm e}^{i \varphi}a_j^\dagger a_i)$, 
namely the magnons acquire a phase\cite{katsura10,onose12}, 
$\varphi={\rm atan}(D/J)$, during propagation, 
which is known to generate a thermal Hall effect\cite{katsura10,onose12,onose10,matsumoto14,lee15}. 
The net phase they carry when going around the closed path is regarded as 
a fictitious magnetic flux that generates the cyclotron motion of magnons. 
As a result, the total momentum of magnon bands is shifted to $-\bm k$ direction, 
which is observed in MnSi\cite{sato2016}. 
\par
The uniform antiferromagnet has two-fold degenerate magnon branches that cross at $\bm k=0$, 
each belonging to sublattice A and B. 
In the main text and in Fig.3{\bf a}-{\bf c} we discussed three cases of antiferromagnets with uniform DM interactions. 
By comparing the ferromagnets and these antiferromagnets, both with uniform DM, 
one finds that the energy bands of antiferromagnets 
are constructed by the combination of ferromagnets; 
The nonreciprocal two bands in Figs.S\ref{fS1}{\bf a} (the same as Figs.3{\bf a} in the main text) 
are constructed by the combination of left and right-shifted ferromagnetic bands carrying quantized magnetic moments 
pointing in the opposite direction. 
By contrast, the DM interactions pointing perpendicular to the magnetic moments 
do not contribute to the shifting of bands in both Figs.S\ref{fS1}{\bf e} and Figs.S\ref{fS1}{\bf b}. 
\par
Notice that for the staggered DM interaction in Fig.S\ref{fS1}{\bf f}, 
the nonreciprocal shift of energy bands does not appear even though $\bm D\parallel \bm M_A , \bm M_B$, 
indicating that the global SIS breaking is required for such phenomena to occur. 
\par
The honeycomb antiferromagnet discussed in the Discussion of the main text 
is also shown in Fig.S\ref{fS1}{\bf g}. 
This system does not break the SIS, so that it is discussed in the context of staggered DM interaction. 
However, since the DM interaction works between next nearest neighbor sites belonging to the same sublattice, 
one can decompose the system into two species of 1D ferromagnets with uniform DM interactions 
pointing in the opposite directions. 
The energy bands indeed show similar structure with the antiferromagnets with uniform DM interaction in Fig.S\ref{fS1}{\bf a}, 
while they accompany the spontaneous toroidal ordering instead of a simple antiferromagnetic order\cite{hayami2016}. 

\begin{figure}[tbp]
\includegraphics[width=8cm]{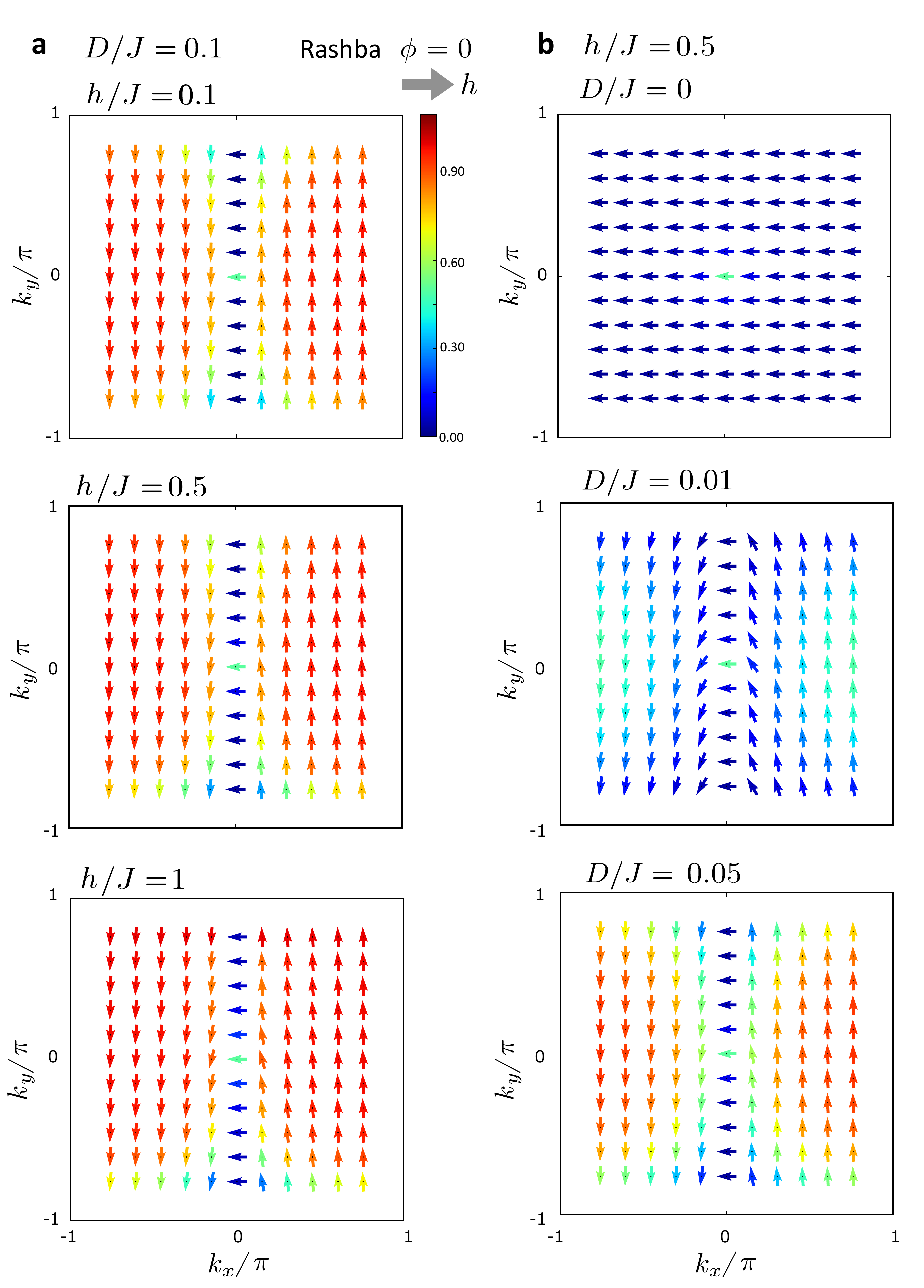}
\caption{ Spin textures of the lower band, $\bm S_{\bm k,-}$, of Rashba magnons at field angle, $\phi=0$, 
for different field strengths ({\bf a}) $h/J=0.1, 0.5$ and 1, 
and DM interaction strengths ({\bf b}) $D/J=0, 0.01$ and 0.05. 
We take $\Lambda=0.05J$. 
}
\label{fS3}
\end{figure}
\subsection{Nonreciprocal magnon bands and spin textures in 2D SIS broken antiferromagnets}
We present the nonreciporocity of energy bands in the 2D case and the spin textures of the upper bands. 
Here, we consider both the in-plane uniform DM interaction $\bm D$, 
and the out-of-plane staggered DM interaction $\bm D_\perp$, shown in Fig.S\ref{fS2}{\bf a}. 
We first examine the magnon band dispersion, which at $\bm D \ne 0$, 
are no longer symmetric about the rotation. 
Figure~S\ref{fS2}{\bf b} shows the one with $\bm h \parallel (\bm e_x+\bm e_y)$ ($\phi=\pi/4$). 
For a given set of $\bm \delta_1$ and $\bm \delta_2$, 
the vectors $\bm D$ along the two bonds point in the directions perpendicular and parallel to $\bm h$, 
respectively. More precisely, the relative relation of $\bm D$ and $\bm h$ 
follow case II and III in Fig.3{\bf d}, respectively, for the two chain directions. 
The cross-sections of bands are shown in Fig.S\ref{fS2}{\bf c}, 
which have the same structure as those found in cases II and III, 
demonstrating full consistency of the idea of decomposing the 2D antiferromagnet into 1D ones. 
The nonreciprocity indeed appears for $\bm k_{\delta_2}$-direction as we saw in case III, 
and not along $\bm k_{\delta_1}$. 
A simple condition to have a nonreciprocity in this system is to have $\theta_M\ne 0$ 
(see Supplementary material B). 
\par
We next show in Fig.S\ref{fS2}{\bf d} and S\ref{fS2}{\bf e}
the spin textures of both the upper and lower bands of both Rashba and Dresselhaus antiferromagnets. 
As explained in the main text, the two bands have spin orientations pointing in the opposite direction 
along the $y$-axis, namely in the direction perpendicular to the field. 
The spin moment parallel to $\bm h$ is the same between the two bands. 
\par
Finally, we show in Fig.S\ref{fS3} the spin textures $\bm S_{\bm k,-}$ of the Rashba magnon with 
different values of field strength and DM interaction, $\bm D$. 
We mentioned in the main text that the spin texture similar to Fig.1{\bf b} (with $h=3J$) 
sustains down to small but finite $h$, as long as we have $\bm D \ne 0$. 
Indeed, we see that the texture is stable against the variation of $h/J=0.1-3$. 
We also find that larger $\bm D$ will give larger variation of $\bm S_{\bm k,-}$, 
but in any case, for $\bm D>0$, the structure is stable. 
For these reasons, we conclude that the spin texture emerges in a wide parameter range. 
